\documentclass[review]{elsarticle}
\usepackage{latexsym}
\usepackage{amsmath,amsthm,amssymb}
\usepackage[latin1]{inputenc}
\usepackage{tikz}
\usepackage[english]{babel}
\usepackage{graphics}
\usepackage{color}
\usepackage[linesnumbered,ruled, vlined, boxed, commentsnumbered]{algorithm2e}
\usepackage[noend]{algorithmic}
\usepackage[unicode, linktocpage=true, bookmarksopen=true, bookmarksopenlevel=0, bookmarksnumbered=true, pdfpagelayout=OneColumn]{hyperref}
\usepackage{paralist}
\usepackage{wrapfig}

\hypersetup{%
pdftitle = {ILP},
pdfsubject = {Information Processing Letters},
pdfkeywords = {ILP, MBrTP, NP-Complete, ...},
pdfstartview = {FitH},
pdfauthor = {H. Kerivin, J. Leblet, G. Simon, F. Zhou}
colorlinks = {false}
}

\newtheorem{thm}{Theorem}
\newtheorem{lem}[thm]{Lemma}
\newcommand{\bpr}{\noindent{\em {Proof\/.} }}
\newcommand{\epr}{\hspace*{\fill}$\Box$\medskip}

\title{Maximum Bounded Rooted-Tree Packing Problem}
\author[Clemson]{Herv\'e Kerivin}
\ead{kerivin@clemson.edu}
\author[lyon]{Jimmy Leblet}
\ead{jimmy.leblet@univ-lyon3.fr}
\author[telecom]{Gwendal Simon}
\ead{gwendal.simon@telecom-bretagne.eu}
\author[telecom]{Fen Zhou}
\ead{fen.zhou@telecom-bretagne@eu}
\address[Clemson]{Department of Mathematical Sciences, Clemson University, South Carolina 29634, USA}
\address[lyon]{\'Ecole Universitaire de Management, Lyon, France}
\address[telecom]{Institut Telecom, Telecom Bretagne, Brest 29238, France}

\begin{document}
\begin{abstract}
  Given a graph and a root, the Maximum Bounded Rooted-Tree Packing
  (MBRTP) problem aims at finding $K$ rooted-trees that span the
  largest subset of vertices, when each vertex has a limited
  outdegree. This problem is motivated by peer-to-peer streaming
  overlays in under-provisioned systems. We prove that the MBRTP
  problem is NP-complete. We present two polynomial-time algorithms
  that computes an optimal solution on complete graphs and trees respectively.
\end{abstract}

\maketitle

\section{Introduction} \label{sec:intro}

Internet is now used to transmit high-definition video streams to
connected TVs and to share user-generated videos captured from
high-quality cameras. Unfortunately, while the demand for transmitting
videos with large bit-rate increases, the available bandwidth for a
connected device has not grown that much (more devices by access
point, awaited technology shift, etc.). Consequences of this lag
between content and infrastructure progresses include that
peer-to-peer (P2P, for short) live streaming systems, as they have
been designed for years, face a new issue: the average upload capacity
of peers is below the stream bit-rate and thus it is physically
impossible to deliver a full-quality service to every
peer~\cite{icc10,streamingcapacity1,layerP2P}. The system is said
under-provisionned.

To address this issue, the use of \emph{multiple description coding}
technique is a promising approach~\cite{layerP2P,mYang2009NCA}. A
video stream is divided into several independent sub-streams,
hereafter called \emph{stripes}, the reception of a subset of stripes
being enough to play the video. The more stripes are received, the
better is the video quality. The main idea developed
in~\cite{layerP2P} is that peers receive more or less degraded data
video but the stream continuity is ensured.

In the current letter, we neglect the practical aspects of P2P systems
(in particular peer churn and incentives to contribute) and we focus
on obtaining a theoretical bound to the problem of delivering multiple
stripes in an under-provisioned overlay network. Each stripe is served
on an independent delivery tree from the source to a subset of
peers. Our objective is to maximize the number of peers spanned in the
multiple trees subject to the upload capacities of peers. The video
quality experienced by a peer depends on the number of trees it
belongs to.


We model the P2P network by an undirected and connected graph $G =
(V,E)$ with $n=|V|$ vertices (or peers) and $m=|E|$ edges. The upload
capacity of vertex $v$, denoted by $c_v$, is the number of stripes
that $v$ can forward. It is a positive integer. The number of distinct
stripes is represented by $K$, which is generally far smaller than
$n$. A specific vertex $r$ in $V$ is given; it is called the {\em
  root} and represents the source peer of the video stream. A {\em
  rooted tree} (or {\em $r$-tree} for short) $T_k = (V_k,E_k)$ is an
acyclic connected subgraph of $G$ with $r \in V_k$. Note that
$(\{r\},\emptyset)$ is the {\em null} $r$-tree. Given a rooted tree
$T$ and a vertex $v$, let $C_T(v)$ be the number of children of $v$ in
$T$. Throughout the paper, we use the shorthand notation $[n]$ to
denote the set $\{1,2,\ldots,n\}$.  A family of $K$ rooted-trees
$\begin{pmatrix}T_1,T_2,\ldots,T_K\end{pmatrix}$ of $G$ is called a
{\em bounded rooted-tree packing} if the following vertex-capacity
requirements are satisfied:
\begin{equation}
 \label{cond:1}
 \sum\limits_{k=1}^K C_{T_k}(v) \leq c_v, \textrm{for all } v \in V.
\end{equation}

The construction of multiple-tree overlay in P2P networks can be generalized as the {\em Maximum Bounded Rooted-Tree Packing (MBRTP) problem}, which consists of finding a bounded rooted-tree packing $\begin{pmatrix} T_k = (V_k,E_k) : k \in [K]\end{pmatrix}$ that spans the maximum number of vertices, that is,  $\sum_{k \in [K]} |V_k|$ is maximized. Whenever only one $r$-tree is sought (i.e., $K=1$), the considered problem will be called the {\em Maximum Bounded Rooted-Tree (MBRT) problem}.

To our knowledge, the MBRTP problem is a new optimization problem. It
is loosely related to the Minimum Bounded Degree Spanning Tree
problem, which tries to determine a minimum-cost spanning tree wherein
any vertex has its degree at most a given value
$k$~\cite{goemans2006mbd}. Variants with non-uniform degree bounds
have also been studied~\cite{KonemannR05}. Differently, the MBRT
problem considers the case where not all vertices can be spanned, the
MBRTP problem extending this formulation to a forest. There is no cost
to minimize here.

In this paper, we prove that both MBRT and MBRTP problems are NP-hard. We present two polynomial-time algorithms that determine the optimal solutions for the MBRTP problem on complete graphs and trees, respectively.

\section{NP-Completeness of the MBRTP problem} \label{sec: np}
We prove the NP-completeness of the MBRTP problem using a
reduction to 3-SAT problem. The decision problem related with MBRTP problem is:

{\sc Question :} Does there exist a bounded $r$-tree packing of size $K$
in which the total number of vertices is greater than or equal to a
positive integer $\Gamma$,
\begin{equation}
\label{eqn: cond2}
i.e. \quad \sum_{k=1}^K |V_k| \geq \Gamma\quad ?
\end{equation}
\begin{thm}
The MBRTP decision problem is NP-complete.
\end{thm}
\bpr We consider the MBRT problem ($K=1$). Verifying that a $r$-tree solves a MBRT instance is polynomial in the size of the problem. Hence the MBRT decision problem
belongs to NP.

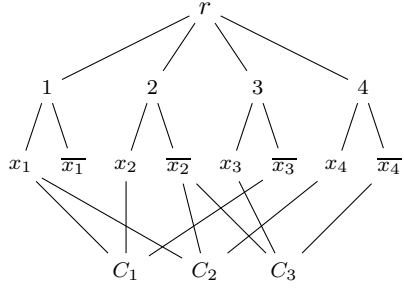
\begin{wrapfigure}{l}{0.5\textwidth}
\begin{center}
\begin{tikzpicture}[scale=0.7]
\node (source) at (0,0) {$r$};

\foreach \i in {1,2,3,4}
{
\node (\i) at (2*\i-5,-1.5) {\footnotesize{$\i$}};
\node (x\i) at (2*\i-5-0.5,-3) {\footnotesize{$x_\i$}};
\node (barx\i) at (2*\i-5+0.5,-3) {\footnotesize{$\overline{x_\i}$}};
\draw (\i) -- (x\i);
\draw (\i) -- (barx\i);
\draw (source) -- (\i);
}

\foreach \j in {1,2,3}
{
\node (c\j) at (1.5*\j-3,-5) {\footnotesize{$C_\j$}};
}

\draw (c1) -- (x1);
\draw (c1) -- (x2);
\draw (c1) -- (barx3);

\draw (c2) -- (x1);
\draw (c2) -- (barx2);
\draw (c2) -- (x4);

\draw (c3) -- (barx2);
\draw (c3) -- (x3);
\draw (c3) -- (barx4);

\end{tikzpicture}
\end{center}
\vspace{-10pt}
\caption{Graph associated with 3-SAT $(x_1 \vee x_2 \vee \overline{x_3}) \wedge (x_1 \vee \overline{x_2} \vee x_4) \wedge (\overline{x_2} \vee x_3 \vee \overline{x_4})$}\label{fig:npcomplete}
\vspace{-5pt}
\end{wrapfigure}

Given an instance of the 3-SAT problem comprising a set of variables $W=\{x_i: 1 \leq i
\leq n\}$ and a set
of clauses $C=\{C_j : 1 \leq j \leq m\}$ on $W$ where $C_j=x_j^1\vee x_j^2\vee x_j^3 $, we define an
instance of the MBRTP problem as follows (see Figure~\ref{fig:npcomplete}).
Let $V'=\{r\}\cup \{i, x_i, \overline{x_i} : 1\leq i \leq n\}\cup \{C_j : 1 \leq j \leq m\}$ and let $E'=\{ri : 1\leq i \leq n\}\cup \{ix_i, i\overline{x_i} : 1\leq i \leq n\} \cup \{ x_j^lC_j: 1\leq j \leq m, 1\leq l\leq 3\}$.
For $1\leq i \leq n$ and for $1\leq j\leq m$, the capacity function is
defined as $c_r=n$, $c_i=1$, $c_{x_i}=c_{\overline{x_i}}=m$ and
$c_{C_j}=0$. Let $\Gamma$ be $1+2n+m$. This MBRT instance can be
constructed in polynomial time in the size of the 3-SAT instance.
 There exists a solution for our MBRT instance if and
only if there exists a truth assignment for $C$.

For the forward implication, assume that there exists a $r$-tree $T=(U,F)$ of $G'=(V',E')$ satisfying (\ref{cond:1}) and (\ref{eqn: cond2}).  Inequalities (\ref{cond:1}) enforce that $U$ cannot contain both $x_i$ and $\overline{x_i}$ for $1\leq i \leq n$, because $c_i=1$ and $c_{C_j}=0$ for $1\leq j \leq m$. Since $|U|\geq\Gamma=1+2n+m$ by (\ref{eqn: cond2}), it follows that
exactly one of $x_i$ and $\overline{x_i}$ belongs to $U$. Since $|U|\geq 1+2n+m$ and $|\{x_i,\overline{x_i}\}\cap U|=1$, every $C_j$ for $1\leq j \leq m$ is in $U$, and $C_j$ is adjacent to exactly one vertex among $\{x_j^1, x_j^2, x_j^3\}$ in $T$. We define the assignment function $\varphi$ as follows: $\varphi(x_i)$ is set to \textit{True} if $x_i\in U$ and \textit{False} if $\overline{x_i}\in U$. We directly obtain that each clause in $C$ has a true value, therefore $\varphi$ is a truth assignment for $C$.

For the backward implication, assume that we have a truth assignment $\varphi'$ for $C$.  We define $W'$ the set of true literals for $\varphi'$, that is $W'=\{x_i : x_i \in W, \varphi'(x_i)=True\}\cup \{\overline{x_i} : x_i \in W, \varphi'(x_i)=False\}$.  We construct the $r$-tree $T'=(U',F')$ of $G'=(V',E')$ as follows. The set of vertices $U'$ is $\{r\} \cup \{i : 1\leq i \leq n\} \cup W' \cup \{C_j : 1 \leq j \leq m\}$. We clearly have $|U'|=1+2n+m$, which makes $T'$ satisfy (\ref{eqn: cond2}).  We leverage on the truth assignment $\varphi'$ to associate with every clause $C_j$ one literal $x'_j$ among $\{x_j^1,x_j^2,x_j^3\} \cap W'$.  The set of edges $F'$ is defined as $\{ri: 1\leq i \leq n\}\cup \{ ix_i : 1\leq i\leq n, x_i \in W'\}\cup \{ i\overline{x_i} : 1\leq i\leq n, \overline{x_i} \in W'\}\cup \{x'_jC_j : 1\leq j \leq m\}$. This construction guarantees that $T'$ is connected and without loss of generality, we can assume that $T'$ is acyclic. Since $r$ belongs to $U'$, $T'$ is a $r$-tree. Moreover, we have that $C_{T'}(r)=n$ and $C_{T'}(C_j)=0$ for $1\leq j \leq m$. Since $C_{T'}(i)=1$, $C_{T'}(x_i) \leq m$, and $C_{T'}(\overline{x_i}) \leq m$  for $1 \leq i \leq n$, we have that $T'$ satisfies~\eqref{cond:1}.

Hence MBRT is NP-complete. Consequently, MBRTP is NP-complete.
\epr

\section{MBRTP on complete graphs} \label{sec:Complete graphs}

We present now a polynomial-time algorithm for complete graphs. Pseudocode
is in Algorithm~\ref{alg:complete-graph}. We prove that it
computes an optimal solution.

This algorithm performs in two stages. The first stage contains $K$
iterative steps. At each step $k \in [K]$, we compute a path rooted at
$r$, noted $s_k$. Let $G_k$ be the complete graph for the set of
vertices having a nonzero capacity at the beginning of the $k$th
step. If the capacity $c_r$ of the root is zero, the path $s_k$ is
empty, otherwise we construct $s_k$ such that it is a \emph{Hamiltonian
  path} on $G_k$. Recall that a Hamiltonian path visits each vertex of
the graph exactly once, and that computing a Hamiltonian path on a
complete graph is trivially in linear-time. Then, we decrease
by one the capacity of every vertex in the path, except the termination
vertex.  The second stage also contains $K$ steps. At each step $k\in
[K]$, we compute a tree $T_k$ for the solution. The tree $T_k$ is
initialized with the path $s_k$, so it is rooted in $r$ and it
contains one branch containing all vertices in $s_k$. Then, every vertex
$v$ in this branch having a nonzero remaining capacity attaches to
$T_k$ vertices that are not yet in $T_k$ until either all vertices in $V$
are in $T_k$, or $v$ has no more capacity.



\begin{algorithm}[!h]
\label{alg:complete-graph}
\SetKwInOut{Input}{Input}
\SetKwInOut{Output}{Output}
\SetKw{KwAnd}{and}
\Input{Complete graph $G$, root $r$, integer $K$, capacity $c_v$ for all $v\in V$}
\Output{$K$ $r$-trees $\begin{pmatrix}T_1,T_2,\ldots,T_K\end{pmatrix}$}
\For{$k \leftarrow 1$ \KwTo $K$}{
      $G_k \leftarrow$ the complete graph of vertex set $\{v|c_v>0,v\in V\}\cup\{r\}$\;
      \lIf{$c_r$ \emph{is zero}}{$s_i \leftarrow (\{r\},\emptyset)$} \Else
          {$s_k \leftarrow$ a Hamiltonian path  in $G_k$ rooted in $r$\;
           \lFor{\emph{every vertex} $v$ \emph{in} $s_k$ \emph{except the termination vertex}}{$c_v \leftarrow c_v -1$}
       }
}
\For{$k \leftarrow 1$ \KwTo $K$}{
    initialize the $r$-tree $T_k = (V_k, E_k)$ from $s_k$\;
    $\overline{V_k} \leftarrow V \setminus V_k$\;
    \For{$v \in V_k$}{
        \While{$\overline{V_{k}}\neq \emptyset$ \KwAnd $c_v>0$}{
          attach a vertex $v' \in \overline{V_k}$ to $r$-tree $T_k$ with $v$ as parent\;
          $\overline{V_k} \leftarrow
          \overline{V_k} \setminus \{v'\}$ ; $c_v \leftarrow c_v - 1$\;
       }
    }
}
\caption{Algorithm for the MBRTP problem in complete graphs}
\end{algorithm}

\begin{thm}
  Given that $G$ is a complete graph, the MBRTP problem can be solved
  in polynomial-time $\mathcal O(nK)$.
\end{thm}
\bpr We prove that Algorithm~\ref{alg:complete-graph} provides the
optimal solution. If $c_r=0$, Algorithm~\ref{alg:complete-graph} results in $K$ null $r-$trees, which is trivially optimal. Here we focus on the case where $c_r> 0$. Since graph $G$ is complete, a complete subgraph can
be built from any subset of vertices of $G$, therefore it is possible
to find a Hamiltonian path for each pruned subgraph $G_k$. The
capacity of the root $r$ decreases by one for every Hamiltonian path
$s_k$ unless either $k$ equals $K$ or $c_r$ is zero. Therefore
$\min\{c_r,K\}$ non-null $r$-trees are produced at the end of the first
stage. In the second stage, for each non-null $r$-tree $T_k$, the
capacity checking process does not finish until any one of the
following conditions is satisfied:
  \begin{itemize}
  \item $\overline{V_k}=\emptyset$, which means that $T_k$ already
    includes all vertices in $G$,
  \item $c_v=0, \forall v \in T_k$, that is, all vertices in $T_k$
    have exhausted their capacity.
  \end{itemize}

  Consequently, the number of spanned vertices in the $\overline k =
  \min\{c_r, K\}$ non-null trees is equal to $\min\big\{\overline
  k+\sum_{v\in V}{c_v},\overline kn \big\}$. In both cases, this number
  reaches the maximum imposed by (\ref{cond:1}) and (\ref{eqn:
    cond2}).

  The first stage takes $\mathcal O(nK)$ time while the latter
  one terminates in time $\mathcal O(nK)$ too. Considering $K\leq n$, Algorithm~\ref{alg:complete-graph} finishes in polynomial time. \epr

\section{MBRTP on trees}

We now consider the case where $G$ is a tree. Designating vertex $r$ as the root, $G$ becomes a $r$-tree. Parameter $K$ is
still the number of trees in the bounded $r$-tree packing.  Given a
peer $v \in V$ and an integer $k\in[K]$, let MBRTP$_k^v$ be a
sub-instance of the MBRTP problem so that the underlying tree is the
subtree of $G$ rooted at $v$ and the number of bounded $r$-trees to
compute is $k$.

First, every vertex $v$ computes the number of spanned vertices in the
optimal solution for every sub-instance MBRTP$^v_k, \textrm{ for all } k \in
[K]$. Each vertex $v$ stores the results in a $K$-dimensional vector
denoted by $g(v)$. The $k$th component of $g(v)$, which is noted
$g(v)_k$, corresponds to the number of spanned vertices counting $v$ itself in the optimal
solution of MBRTP$^v_k$. Obviously, $g(v)_k$ is monotonically increasing with respect to $k$ for any vertex $v$. For a leaf $v$ of $G$, the solution of MBRTP$^v_k$ is $k$ times the vertex $v$ itself (formally $g(v)_k = k$).


A non-leaf vertex $u$ leverages on the computations that have been made
by its children to compute its own vector $g(u)$. We define $C_G(u)$
as the set of children of $u$ in $G$, and $n_G(u) = |C_G(u)|$. Let $x_{uv}^i,
i\in[k], u\in V, v \in C_G(u)$ be a set of binary variables where
$x_{uv}^i$ equals 1 if $u$ allocates exactly $i$ capacities to its
child $v$ (\textit{i.e.}, $v$ is the child of $u$ in $i$ of the $k$ bounded
$r$-trees), otherwise it is 0. If the variable $x_{uv}^i$ is 1, then
$v$ is served with $i$ stripes that $v$ is able to relay in the
sub-trees rooted at itself, spanning exactly $g(v)_i$ peers counting $v$ itself. When all variables $x_{uv}^i, i\in [k]$ are zero for $v$, vertex $v$ will receive zero stripe and neither $v$ nor its children will be spanned in the $k$ bounded $r$-trees. Given an integer $k \in [K]$, the capacity $c_u$ of vertex $u$, and
a vector $g(v)$ of every vertex $v \in C_G(u)$, the value of $g(u)_k$
can be obtained by solving the following Non-Standard Multiple-Choice
Knapsack Problem (NS-MCKP):
\begin{align}
\notag  \max \qquad & k+\sum_{v \in C_{G(u)}}\sum_{i=1}^{k}g(v)_i\times x_{uv}^i \quad \textrm{(NS-MCKP)}\\
\label{eq:nomorecap}
\textrm{s.t.} & \quad \sum_{v \in C_{G}(u)} \sum_{i=1}^{k}i\times x_{uv}^i \leq c_u,\\
\label{eq:nomorestripe}& \quad \sum_{i=1}^k x_{uv}^i \leq 1 \quad \forall v \in C_{G}(u),\\
\notag              &\quad x_{uv}^i \in \{0, 1\} \quad \forall v \in C_{G}(u) \quad \forall i \in [k].
\end{align}
Constraints~(\ref{eq:nomorecap}) ensure that the capacity constraint
of $u$ should not be violated by the sum of the capacity allocated to
its children. Constraints~(\ref{eq:nomorestripe}) mean that the number
of stripes sent by $u$ to $v$ is in $\{0,1,2,...,k\}$.

\begin{lem}
The above NS-MCKP can be solved in time $\mathcal O(k^2\times n^2_G(u))$.
\end{lem}
\bpr We use dynamic programming as follows.  Without loss of
generality, we label the children of $u$ from $v_1$ to
$v_{n_G(u)}$. Given two integers $d\in [n_G(u)]$ and $c \in
\{0\}\cup[c_u]$, let NS-MCKP$_d^k(c)$ be the sub-instance of NS-MCKP where the set of children of $u$ is restricted to $\{v_1,...,v_d\}$, $k$ is the number of received stripes,
and the capacity is $c$.  We denote by $f_d^k(c)$  the optimal solution
value for the NS-MCKP$_d^k(c)$. When $d=1$ we have
\begin{align}
\notag f_1^k(c) = \left\{
\begin{array}{ll}
\notag k     &  \textrm{ if } c=0,\\
\notag k+ g(v_1)_c &        \textrm{ if } 1 \leq c < k,\\
\notag k+ g(v_1)_k &  \textrm{ if } k \leq c.
\end{array}
\right.
\end{align}
When $2\leq d \leq n_G(u)$, $f_d^k(0)$ is $k$, and whenever $c$ is greater or equal to $k
\times d$, then $f_d^k(c) = k+\sum_{i\in [d]} g(v_i)_k$, that is, the optimal solution consists of assigning $k$
capacities to every child in $\{v_1,...,v_d\}$. For any value of $c$ ranging from $1$ to
$k\times d$, the solution $f_d^k(c)$ is computed by comparing
$f_{d-1}^k(c)$ with what can be obtained if $u$ decides to allocate $i
\in [k]$, capacities to the vertex $v_d$. Formally,
\begin{align}
\notag f_d^k(c) = \left\{
\begin{array}{ll}
\notag k     &  \textrm{ if } c=0,\\
\notag \max \big\{f_{d-1}^k(c),\max \{f_{d-1}^k(c-i)+g(v_d)_i: i \in [k]\}\big\}
                      &        \textrm{ if } 1 \leq c < k\times d,\\
\notag k+\sum_{i\in [d]} g(v_i)_k &  \textrm{ if } k\times d \leq c.
\end{array}
\right.
\end{align}

The solution of the NS-MCKP problem is $f_{n_G(u)}^{k}(c_u)$. For each
value of $d$, the computation requires at most $\mathcal O(k \times
\min\{c_u, k\times d\})$ comparisons, as a result the overall time
complexity of solving the NS-MCKP problem is $\mathcal O(k^2\times n^2_G(u))$.
\epr

\begin{thm}
Given that $G$ is a tree, the MBRTP problem can be solved in polynomial-time $\mathcal O(n^3K^3)$.
\end{thm}
\bpr
 A vertex $u$ computes its vector $g(u)$ with $g(u)_k = f_{n_G(u)}^k(c_u)$
for any $k \in [K]$. This requires solving the NS-MCKP problem $K$
times. The value of $g(r)_K=f_{n_G(r)}^K(c_r)$ corresponds to the
optimal solution of the MBRTP problem as it corresponds to the optimal
solution of MBRTP$^r_K$. It should be noted that $f_{n_G(r)}^K(c_r)$
cannot be computed before knowing all vectors $g(.)$ of the root's
children. Consequently, the computation of vector $g(.)$ should be
done from the leaves to the root $r$ in a breadth-first manner, which
requires solving the NS-MCKP problem $nK$ times in total. We have $n_G(u) < n$, thus the overall time complexity of the proposed algorithm is $\mathcal O(n^3K^3)$ provided that $G$ is a tree. As $K\leq n$, it is polynomial. \epr

\section{Conclusion} \label{sec:conclusion}
In this paper we investigate the Maximum Bounded Rooted-Tree Packing Problem in under-provisioned P2P networks, which aims at maximizing the number of peers that are spanned in the multiple video delivery trees under the capacity constraint of peers. We prove that the MBRTP problem is NP-hard, while it can be polynomially solved on both complete graphs and trees.


%
%
%
%


\bibliographystyle{plain}
\bibliography{related}
\end{document}